\documentclass[12pt]{iopart}
\usepackage{slashed}

\def\nc{\newcommand}
\nc{\be}{\begin{equation}}
\nc{\ee}{\end{equation}}
\nc{\bea}{\begin{eqnarray}}
\nc{\eea}{\end{eqnarray}}
\def\nn{\nonumber}

\nc{\el}{\mbox{\boldmath $l$}}
\nc{\en}{\mbox{\boldmath $n$}}
\nc{\emm}{\mbox{\boldmath $m$}}
\nc{\emmB}{\mbox{\boldmath $\overline{m}$}}
\nc{\emP}{\emm+\emmB}
\nc{\emM}{\emm-\emmB}
\nc{\cR}{\mathcal{R}}
\nc{\cQ}{\mathcal{Q}}
\nc{\cP}{\mathcal{P}}
\nc{\xd}{\mathbf d}
\nc{\rt}{\sqrt{2}}

\nc{\alphaB}{\overline{\alpha}}
\nc{\betaB}{\overline{\beta}}
\nc{\gammaB}{\overline{\gamma}}
\nc{\deltaB}{\overline{\delta}}
\nc{\epsilonB}{\overline{\epsilon}}
\nc{\kappaB}{\overline{\kappa}}
\nc{\lambdaB}{\overline{\lambda}}
\nc{\muB}{\overline{\mu}}
\nc{\nuB}{\overline{\nu}}
\nc{\rhoB}{\overline{\rho}}
\nc{\sigmaB}{\overline{\sigma}}
\nc{\tauB}{\overline{\tau}}
\nc{\piB}{\overline{\pi}}

\nc{\pd}{\partial}
\nc{\pdR}{\pd_\cR}
\nc{\ddx}[1]{\frac{\pd}{\pd x^{#1}}}
\nc{\dx}[1]{\xd x^{#1}}

\nc{\refeq}[1]{(\ref{#1})}

\begin{document}
\jl{6}

\title[One-Killing-vector canonical-frame equations]{Vacuum spacetimes with a spacelike, 
hypersurface-orthogonal Killing vector: \\ reduced equations in a canonical frame}

\author{S. Bonanos$\,\dag$}

\address{\dag\ Institute of Nuclear Physics, NCSR Demokritos,\\
15310 Aghia Paraskevi, Attiki, Greece}

\begin{abstract}

The Newman-Penrose equations for spacetimes having one spacelike Killing vector 
are reduced --  in a geometrically defined ``canonical frame'' --   to a 
minimal set, and its differential structure is studied.  
Expressions for the frame vectors in  an arbitrary coordinate basis
are given, and coordinate-independent choices of the metric functions are suggested which make the 
components of the Ricci tensor in the direction of the  Killing vector vanish.

\end{abstract}


\section{Introduction}
Interest in the vacuum Einstein equations with a non-null Killing vector 
(KV) has been 
rekindled recently by the discovery of Fayos and Sopuerta \cite{FS01} that
the Weyl scalars in the Newman-Penrose (NP) formalism can be expressed 
algebraically in terms of
the spin coefficients and the norm of the Killing vector. Thus, 
Steele \cite{STL02} has 
extended the results of Fayos and Sopuerta to the case of a proper homothetic
vector, while Ludwig \cite{LUD02}, using the
GHP formalism, further extended these results to the non-vacuum case and
showed how the formalism can be used to determine all conformal Killing vectors
of a particular non-vacuum metric. 

The formalisms developed in these papers are completely general. The 
components of the Killing vector and of the ``Papapetrou field'' 
\cite{FS99} -- the exterior derivative of the Killing 1-form -- 
are written down without making any gauge 
choices (except in \cite{LUD02}, where a  "preferred tetrad"  similar to the one 
defined here, is used),
and the integrability conditions relating these quantities are found.  
As a result, these formalisms 
involve a number of variables that can be chosen arbitrarily (to fix the frame 
relative to the Killing vector), resulting in a redundant set of equations. 
Moreover, due to the complexity of these equations, the implications of the 
vacuum Bianchi identities (when the Weyl scalars are substituted for in 
terms of the spin coefficients) are not considered in any of the 
works mentioned above. Thus, despite the great number of equations contained
in these papers, 
the entire system of equations to be solved has yet to be written down.

In this paper we examine the particular case of a vacuum spacetime with a 
spacelike, hypersurface-orthogonal Killing vector, whose norm has a 
spacelike gradient. These assumptions hold true, at least near 
infinity, for asymptotically flat axisymmetric spacetimes. Thus, they 
are appropriate for studying the simplest 
physical problem where the full dynamical content of 
General Relativity can be manifested: the coalescence of two Schwarzschild 
black holes falling toward one another along the line joining their 
centers. 
Our choice of notation will refer to this axisymmetric problem, even though 
the equations will be valid in any spacetime having the assumed 
properties.

Our main result is that, 
by making a natural choice of frame based on the assumed properties 
of the KV, we are able to show that the entire set 
of Ricci and Bianchi equations in the NP formalism  reduce to a set of 
17 real equations (11 for vacuum). We also recover,  in a 
coordinate-free way, Waylen's 
\cite{WAYL87} result  that, when the ``main'' equations are 
satisfied, the contracted Bianchi identities can be formally 
integrated in terms of a function satisfying the wave equation.

In section 2  we define the canonical frame (up to a boost) and deduce 
relationships among the spin coefficients that follow from our choice 
of frame, while in section 3 we obtain 
the reduced set of equations and elucidate their structure.
 In section 4 we give general coordinate expressions 
for the frame components which make the Ricci tensor have vanishing 
components in the direction of the KV. Finally, in section 5 we discuss ways of  
choosing the remaining frame and coordinate freedom, 
as well as additional conditions that can be imposed  in order to seek solutions 
satisfying extra physical or mathematical requirements.
 
\section{The Canonical Frame and the Papapetrou Field}\label{secFR}
The NP equations are invariant under arbitrary frame rotations
(arbitrary Lorentz transformations). We can use this six-real-parameter freedom
to fix the frame relative to the geometric structures assumed for the 
spacetime. 

First, we will use three parameters to rotate our frame so that the space-like 
Killing vector $\xi$ points in the direction of ($\emM$). 
(We use the standard NP notation \cite{PandR}: the 
complex null-tetrad basis $\{\el, \en, \emm, \emmB \}$ is normalized to 
$\el \cdot\en = -\emm\cdot\emmB= 1$.) If the (closed) Killing trajectories are 
parametrized by the parameter $\varphi$, then the assumption that the KV is 
hypersurface orthogonal implies that we can write
\bea
\xi^{a} \ddx{a}  = \frac{\pd}{\pd \varphi} = \cR 
\frac{(m^{a}-\overline{m}^{a})}{\rmi \,\rt} \ddx{a},  \mbox{ 
and  } 
\label{eq:dksiUP}\\ \xi_{a}\dx{a} = -\cR^2 
\xd \varphi = \cR \frac{(m_{a}-\overline{m}_{a})}{\rmi \,\rt}\dx{a},
\label{eq:dphi}
\eea
where $\cR$ is a scalar function giving the norm of the KV:
\be
\xi^{a} \xi_{a} =-\cR^2.
\ee
The 3-surfaces orthogonal to the Killing trajectories are spanned by 
the vectors $\el, \:\en, \:(\emP)$. In the following, we will 
restrict the symbols $\{\el, \:\en, \:\emm, \:\emmB \}$ to denote the 
corresponding co-vectors (differential forms) while the 
vectors (differential operators) will be denoted by the standard symbols 
$\{D, \:\Delta, \:\delta, \:\deltaB \}$. Thus the exterior derivative of an 
arbitrary function $f$ will be written
\be
\xd f = \el \,\Delta f+\en \,D f-\emm \,\deltaB f-\emmB \,\delta f.
\ee

The choice of frame  $\emm-\emmB\:\sim\,$ to a hypersurface orthogonal Killing vector 
drastically simplifies the NP equations: all spin coefficients 
become real (see the Appendix), while $\deltaB f= \delta f$ on any function 
$f$ of the 3 essential coordinates. Thus we can drop the bar from 
$\deltaB$ (and, of course, from the spin coefficients) in all equations.

Next, we note that the gradient of the norm of the KV will have components in the 
$\{\el, \:\en, \:\emP \}$ directions only and, by assumption, 
will be spacelike. We can, therefore, use two degrees of 
freedom to further restrict our frame by requiring that   
\be \xd \cR  \sim \emP \hspace{25pt} \Longleftrightarrow \hspace{25pt}  
D\cR=0=\Delta\cR. \label{eq:dR}
\ee
The frame is now 
completely determined up to a boost in the $\{\el, \:\en\}$ plane 
parametrized by an arbitrary scalar function $A$: 
\be
\el\rightarrow A\:\el, \hspace{40pt}\en\rightarrow \frac{1}{A}\en. 
\label{BoostA}
\ee
There does not seem to be a ``natural'' choice for eliminating this 
freedom, 
based on the assumed properties of the Killing vector. We will 
discuss possible choices of this last gauge degree of freedom in 
section \ref{secBOO}.

Now, evaluating\footnote{The exterior derivatives of 
the coframe 1-forms are given in the Appendix.} $\xd [(\emM)/\cR]=0$, which follows from equation \refeq{eq:dphi}
with all spin coefficients real and using equation \refeq{eq:dR}, we find
\be
\lambda=\mu,\hspace{25pt} \sigma=\rho,\hspace{25pt} 
\delta\,\cR=-(\alpha-\beta)\cR, \label{eq:ddph}
\ee
which determines the proportionality factor in \refeq{eq:dR} so that the 
complex co-vector $\emm\,$  takes the form
\be
\emm =  \frac{\xd \cR}{2(\alpha-\beta)\cR}-\frac{\rmi}{\rt} \cR \,\xd\,\varphi. \label{eq:EM}
\ee 
Next, evaluating the condition $\xd \xd \cR =0$ using equations 
\refeq{eq:dR} and \refeq{eq:ddph}, we obtain
\bea
D(\alpha -\beta)& = & 2\,\rho\,(\alpha -\beta) 
 \label{eq:DQ}  \\
 \Delta(\alpha -\beta) & = & -\,2\mu\,(\alpha -\beta)
 \label{eq:DelQ}  \\
 \hspace{40pt}\pi & = & -\tau.
 \label{eq:pt}
\eea
At this point it is convenient to introduce another scalar function, 
$\cQ$, 
by the equation
\be
(\alpha-\beta)\cR=-\cQ/\rt, \hspace{40pt}\mbox{so that } \:
\cR^{,a}\cR_{,a}=-\cQ^2.\label{Qdef}
\ee
The function $\cQ$ takes the value of unity at infinity (asymptotic 
flatness), and on the axis ($\cR=0$) when the coordinate $\varphi$ has the standard 
periodicity of $2\pi$ (``regularity condition'' -- see  \cite{KRAM}). 
In terms of $\cQ$, equations \refeq{eq:DQ}, \refeq{eq:DelQ} take the form
\be
D \cQ=2\,\rho\,\cQ , \hspace{30pt} \Delta \cQ=-2\,\mu\,\cQ. \label{eq:D12Q}
\ee

\subsection{The Papapetrou Field and the Equation $R_{ab}\xi^{b}=0$}
Using equations \refeq{eq:dphi} and \refeq{eq:EM}, the Papapetrou field in this 
frame takes the form
\be
F=\xd(\xi_{a}\dx{a})=-2\,\cR\,\xd\cR\wedge\xd\varphi=-2\,\rmi\,\cQ\,\emm\wedge\emmB. 
\label{eq:Fpap}
\ee
Now, by virtue of the identity 
$\xi_{b;c;d}-\xi_{b;d;c}=\xi^{a}R_{abcd}$ and Killing's equations 
$\xi_{a;b}+\xi_{b;a}=0$, the Papapetrou field satisfies the equation 
$F^{ab}{}_{;b}=-2\,\xi^{a;b}{}_{;b}=2\,R^{a}{}_{b}\xi^{b}$, so that in vacuum 
(or when  $R^{a}{}_{b}\xi^{b}$ vanishes), the Hodge-dual of 
F\footnote{Note that the self-dual part of $F$, $F-\rmi 
^{*}F, \mbox{ equals } 2\,\rmi\,\cQ\,(\el\wedge\en-\emm\wedge\emmB)$ so the ``canonical frame'' 
defined in this paper based on the properties of the KV coincides with the 
``preferred frame'' (when $F$ is non-null) defined in 
references \cite{FS01, LUD02} based on the coincidence of the null 
eigendirections of 
the Papapetrou field with the NP frame vectors $\el,\,\en$.} is closed,
\be
\xd(^{*}F)=\xd(-2 \,\cQ \,\el\wedge\en)=0. \label{eq:starFpap}
\ee
This gives the single equation
\be
\delta\cQ=2\,\tau\,\cQ. \label{eq:d3Q}
\ee
Finally, using  equations \refeq{eq:D12Q} and \refeq{eq:d3Q}, the 
integrability condition $\xd \xd \cQ =0$ gives
\bea
D\tau-\delta\rho=\kappa\, \mu-\rho\,(\alpha+\beta-\tau), \label{Deetau}\\
D\mu+\Delta\rho=2\,(\rho\,\gamma-\mu\,\epsilon), \label{DeemDelr}\\
\Delta\tau+\delta\mu=\nu\,\rho-\mu\,(\alpha+\beta+\tau). \label{Deltau}
\eea

In summary, we restrict the choice of frame by imposing 
the five conditions:
(i) the components of $\xi$ in the $\el, \,\en, \, (\emP)$ directions 
vanish, and  
(ii) the gradient of the norm of the KV points in the direction of 
$\emP$, which, for a hypersurface-orthogonal KV, implies 
that the spin coefficients satisfy $\sigma=\rho,\mbox{ and }\lambda=\mu$.
 Since $\emP$ is now hypersurface orthogonal,  the condition 
 $\pi+\tau=0$ follows.
 The remaining equations obtained in this section are the result of applying 
 the commutators to the functions $\cR, \,\cQ$  and 
 requiring $R_{ab}\xi^{a}\xi^{b}$ to vanish. Being 
 integrability conditions, they are contained in the Ricci and Bianchi identities 
 as will be shown in the next section.
 
\section{The Reduced NP Equations}\label{redEQU}
We now examine the Ricci identities with 
$\sigma=\rho,\:\lambda=\mu,\:\pi=-\tau\mbox{ and }\deltaB=\delta$ and 
assuming all spin coefficients real. As these equations give real
expressions for $\Psi_{a},\: \Phi_{ab}$ we can assume 
$\Phi_{ba}=\overline{\Phi_{ab}}=\Phi_{ab}$. 
Then the 18 Ricci equations can be combined to give 
\begin{itemize}
  \item  Six equations defining the 5 $\Psi_{a}$  
  (the expression for $\Psi_{2}$ can be obtained in two ways)
  \bea
  \Psi_{0}=-2\,\kappa\,(\alpha-\beta)+\Phi_{00}, \nn\\
  \Psi_{1}=-2\,\rho\,(\alpha-\beta)+\Phi_{01},\nn \\
  \Psi_{2}=-2\,\tau\,(\alpha-\beta)+\Phi_{02}-2\Lambda, \label{PSIvalues}\\
  \Psi_{3}=2\,\mu\,(\alpha-\beta)+\Phi_{12}, \nn\\
  \Psi_{4}=2\,\nu\,(\alpha-\beta)+\Phi_{22}.\nn
  \eea
    \item  Five ``coframe integrability conditions'', obtained by exterior 
    differentiating
  equations \refeq{finXDel} - \refeq{finXDemM} in the Appendix. These 
  equations, being independent of the Weyl and Ricci tensor 
  components, are called 
  ``eliminant relations'' by Chandrasekhar \cite{CHANDRA}. They are best 
  written using the  ``divergence operators'' $\slashed{D},\; 
  \slashed{\Delta},\:\slashed{\delta}\:$ introduced in \cite{SB96}:
  \be
    \fl\slashed{D}\equiv D+\epsilon+\epsilonB-\rho-\rhoB,\hspace{15pt}
     \slashed{\Delta}\equiv \Delta-\gamma-\gammaB+\mu+\muB,\hspace{15pt}
     \slashed{\delta}\equiv\delta+\beta-\alphaB+\piB-\tau.\label{slashDEF}
     \ee
     Using these operators, which appear naturally in the exterior differentiation of 
     3-forms, the coframe integrability conditions read:
    \bea
   \slashed{D}(\tau-2\,\beta)- 
   \slashed{\Delta}\kappa+2\,\slashed{\delta}\epsilon=0,\label{ELIM1}\\
   \slashed{D}\mu+\slashed{\Delta}\rho=0,\label{DmDr}\\
   \slashed{D}\nu+
   \slashed{\Delta}(\tau+2\,\alpha)-2\,\slashed{\delta}\gamma=0,\\
   \slashed{D}(\alpha-\beta)=2\,\epsilon(\alpha-\beta),\label{Deeab}\\
   \slashed{\Delta}(\alpha-\beta)=-2\,\gamma(\alpha-\beta).\label{Delab}
    \eea
   Three of these equations have already been obtained in computing 
   the integrability conditions of $\xd\cR,\;\xd\cQ\:$: Equations 
   \refeq{Deeab}, \refeq{Delab} are the same as equations \refeq{eq:DQ} and 
   \refeq{eq:DelQ}, 
   while equation \refeq{DmDr} is the same as equation \refeq{DeemDelr}.
    \item  Seven remaining equations determining the Ricci tensor. 
    Using the abbreviations 
 \bea
 G_{1}=(\epsilon-\rho)(\tau+\alpha-\beta)-\kappa\,(\mu+\gamma), \\
 G_{0}=\gamma\,\rho+\epsilon\,\mu, \label{G0}\\
 G_{-1}=(\mu-\gamma)(\tau+\alpha-\beta)-\nu\,(\rho+\epsilon),\\
 \Sigma_{0}=\tau^{2}+\kappa\,\nu+2\,\tau\,(\alpha-\beta),
    \eea
 they take the form
 \bea
\slashed{D}\rho- 
\slashed{\delta}\kappa=4\,\epsilon\,\rho-2\,\kappa\,(\alpha+\beta)+\Phi_{00},\label{ric1}\\
\slashed{D}(\alpha+\beta)-2\, 
\slashed{\delta}\epsilon=2\,(G_{1}+\Phi_{01}),\\
\slashed{D}\mu- \slashed{\Delta}\rho+2\, 
\slashed{\delta}\tau=2\,(-\Sigma_{0}+\Phi_{02}),\\
\slashed{D}\gamma- 
\slashed{\Delta}\epsilon=-\Sigma_{0}-2\,G_{0}+\Phi_{02}+\Phi_{11}-3\,\Lambda,\\
\slashed{\delta}(\alpha-\beta)=-\Phi_{02}+\Phi_{11}+3\,\Lambda,\label{dstrF}\\
  2\, \slashed{\delta}\gamma- 
  \slashed{\Delta}(\alpha+\beta)=2\,(G_{-1}+\Phi_{12}),\\
   \slashed{\delta}\nu- \slashed{\Delta}\mu=4\,\gamma\,\mu-2\,\nu\, 
   (\alpha+\beta)+\Phi_{22}.\label{ric7}
  \eea
  \end{itemize}
 Using the definition of $\cQ$ (equation \refeq{Qdef}) and \refeq{eq:ddph}, 
 we find that equation \refeq{dstrF} is the same as \refeq{eq:d3Q} when 
  \be
  \Phi_{11}+3\,\Lambda-\Phi_{02} = \frac{1}{2}R_{ab}\xi^{a}\xi^{b}=0.\label{R44eq}
  \ee

  \subsection{The Bianchi Identities}
  When the values of $\Psi_{a}$ given by \refeq{PSIvalues} together with 
  $\sigma=\rho,\:\lambda=\mu,\:\pi=-\tau\mbox{ and }\deltaB=\delta$ are 
  substituted into the Bianchi identities, and the Ricci identities 
  with $\Phi_{ba}=\Phi_{ab}$ are used,
  one finds, after a long calculation,  that they can be reduced to five independent equations: the
  three contracted Bianchi identities, involving the components of the 
  Ricci tensor only,
 \bea
\fl\hspace{25pt}D(\Phi_{11}+3\,\Lambda)-2\,\delta\Phi_{01}+\Delta\Phi_{00} =
 2\,(2\,\gamma-\mu)\Phi_{00}-2\,(2\,\alpha+3\,\tau)\Phi_{01}\nn\\
\hspace{110pt} +2\,\rho\,(\Phi_{02}+2\,\Phi_{11})-2\,\kappa\,\Phi_{12},\\
\fl D\Phi_{12}-\delta(\Phi_{02}+\Phi_{11}-3\,\Lambda)+\Delta\Phi_{01} 
=\nu\,\Phi_{00}+2\,(\gamma-2\,\mu)\Phi_{01}-2\,(\alpha-\beta+\tau)\Phi_{02}\nn\\
\hspace{110pt}-4\,\tau\,\Phi_{11}+2\,(2\,\rho-\epsilon)\Phi_{12}-\kappa\,\Phi_{22},\\
\fl \hspace{25pt}D\Phi_{22}-2\,\delta\Phi_{12}+\Delta(\Phi_{11}+3\,\Lambda)
=2\,\nu\,\Phi_{01}-2\,\mu\,(\Phi_{02}+2\Phi_{11})\nn\\
\hspace{110pt}+2\,(2\,\beta-3\,\tau)\Phi_{12}+2\,(\rho-2\,\epsilon)\Phi_{22},
  \eea
  and the following two  equations
  \bea
  \fl (D-4\rho)(\Phi_{11}+3\,\Lambda-\Phi_{02})
  +2\,(\alpha-\beta)(D\tau-\delta\rho+\rho\,(\alpha+\beta-\tau)-\kappa\, \mu)=0,\\
  \fl (\Delta+4\mu)(\Phi_{11}+3\,\Lambda-\Phi_{02})
  +2\,(\alpha-\beta)(\Delta\tau+\delta\mu+\mu\,(\alpha+\beta+\tau)-\nu\,\rho)=0.
    \eea
  These last two equations 
  reduce to equations \refeq{Deetau}, \refeq{Deltau} when \refeq{R44eq} 
  holds. This was to be expected since equations \refeq{Deetau}-\refeq{Deltau} 
  were derived using  \refeq{eq:d3Q}, which is equation \refeq{dstrF} with 
  vanishing rhs. Collecting everything together we conclude that, in 
  this frame, the complete\footnote{Apart from  
  three equations obtained by letting the commutators 
  \refeq{commut} act on a  function independent of $\cR,\,\cQ$.} set of  NP equations 
  for spacetimes with  one hypersurface-orthogonal KV,  are given by 
  the 17 equations in this section (12 Ricci and 5 Bianchi). The Weyl 
  scalars are given by \refeq{PSIvalues} and the Ricci 
  tensor identically  satifies $\xi_{[a}R_{b]c}\xi^{c}=0\: (\: 
  \leftrightarrow\:\Phi_{ba}=\Phi_{ab})$.
  
 In  section \ref{secCOORD}, we show that  two of these equations (\ref{Deeab}, \ref{Delab}) 
  can be satisfied identically by using  equation \refeq{eq:D12Q}  to \emph{define}  $\rho \,\en-\mu \,\el$.
  When the Ricci tensor satisfies $ R_{ab}\xi^{a}\xi^{b}=0$, 
  one more equation (\ref{dstrF}) can be reduced to an identity by  choosing 
  certain  metric functions appropriately. Finally, in  vacuum, 
  the three contracted Bianchi identities disappear, so we are left 
  with 11 (real)  equations.
 
\subsection{Main and Subsidiary Equations}
Ever since the classic paper of Bondi \emph{et al.} \cite{Bondi62}, it 
is instructive
to split the seven  Ricci equations (\ref{ric1}-\ref{ric7})  into 4 ``main'' and
3 ``subsidiary'' equations, such that, when the ``main'' equations are 
satisfied everywhere, the contracted Bianchi identities impose extra conditions 
on an ``initial'' hypersurface. In a paper that has received less 
attention than it deserves\footnote{This paper gives a prescription 
for constructing a power series  solution of the axisymmetric vacuum Einstein equations, valid 
near the axis, and depending on an arbitrary function of two 
variables --  much as \cite{Bondi62} gives the series solution near 
future null infinity in terms of an arbitrary ``news'' function $c(u,\theta)$. However, unlike \cite{Bondi62} where four main 
equations need to be satisfied, in \cite{WAYL87} one main equation is 
satisfied identically.}, Waylen \cite{WAYL87} has studied the 
axisymmetric vacuum Einstein equations. Using a particular 
coordinate system (one coordinate being the scalar function $\cR$) in 
which \refeq{R44eq} is satisfied identically,
he has shown that
\begin{enumerate}
   \item  Near the symmetry axis, analytic (power series) solutions 
depending on an arbitrary function of two variables can be constructed 
by solving the ``main'' equations.
    \item  Two of the  Bianchi identities can then be
 satisfied identically by letting the three  remaining components of the Ricci 
tensor  be proportional to the derivatives of a single scalar 
function $\cP$.
    \item  The third Bianchi identity then implies that the function 
    $\cP$ satisfies the wave equation;  and the only solution of that 
    equation that
    behaves well on the axis is the trivial solution  $\cP=0$.
\end{enumerate}

Remarkably, Waylen's integration of the Bianchi identities can be 
carried out in our formalism 
\emph{without resorting to a particular coordinate system}. 
Following Waylen, let us take as main equations the set 
$\{\Phi_{00}=0,\:\Phi_{11}=3\,\Lambda,\:\Phi_{22}=0\}$ together with  
equation \refeq{R44eq}. 
 Then, expressing the non-vanishing Ricci components in terms of 
$\Phi_{01},\,\Phi_{11},\,\Phi_{12}$, the Bianchi identities become
\bea
D \Phi_{11} - \delta \Phi_{01} + 
  (2\,\alpha  + 3\,\tau )\,\Phi_{01} - 
  4\,\rho \,\Phi_{11} + \kappa \,\Phi_{12}=0,\\
\fl D \Phi_{12} - 2\,\delta \Phi_{11} + 
  \Delta \Phi_{01} - 2\,
  (\gamma  - 2\,\mu )\,\Phi_{01} + 
     4\,(\alpha  -\beta  + 
     2\,\tau )\Phi_{11}+2\,( \epsilon -
     2\,\rho )\,\Phi_{12} =0,\\
\delta \Phi_{12} - \Delta \Phi_{11} + 
  \nu \,\Phi_{01} - 4\,\mu \,\Phi_{11} + 
  (2\,\beta \ - 3\,\tau )\,\Phi_{12}=0.
\eea
Now, the first and last of these equations are satisfied identically, by virtue 
of the commutators \refeq{commut} and the known derivatives of $\cQ$ 
and $\cR$,  if we set
\be
\Phi_{01}=\frac{\cQ}{\cR}\,D\cP,\hspace{30pt}\Phi_{11}=\frac{\cQ}{\cR}\,\delta\cP,
\hspace{30pt}\Phi_{12}=\frac{\cQ}{\cR}\,\Delta\cP,
\ee
for some scalar function $\cP$. The second equation (apart from an overall factor $\cQ/\cR$) then becomes
\be
\slashed{\Delta}(D\cP)+ \slashed{D}(\Delta \cP)-2 
 \slashed{\delta}(\delta\cP)=0,
\ee
which is the wave equation for $\cP$, the left-hand-side being equal 
to $ -^{*}\xd(^{*}\xd\cP)$.

The ``main'' equations to be solved are then 
\bea
\slashed{D}\rho- 
\slashed{\delta}\kappa=4\,\epsilon\,\rho-2\,\kappa\,(\alpha+\beta),\label{main1}\\
\slashed{D}(\mu-2\,\gamma)- 
\slashed{\Delta}(\rho-2\,\epsilon)+2\, \slashed{\delta}\tau=4\,G_{0},\label{main2}\\
\slashed{\delta}\nu- \slashed{\Delta}\mu=4\,\gamma\,\mu-2\,\nu\, (\alpha+\beta),\label{main3}
  \eea
together with the ``integrability conditions'' -- equations 
\refeq{Deetau} - \refeq{Deltau} and \refeq{ELIM1} - 
\refeq{Delab}.\footnote{ These integrability conditions are identities when 
the spin coefficients are expressed in terms of the derivatives of the 
components of the frame vectors in a coordinate frame.} And 
Waylen's result suggests that a solution of these equations that is 
 well-behaved near the axis $\cR=0$ will also satisfy 
the remaining three Ricci equations.
\section{General Coordinate Expressions}\label{secCOORD}
Let $ x^{a}\:( a=1,2,3)\:$ be the coordinate labels of an arbitrary coordinate system on the 
3-surfaces orthogonal to the Killing trajectories. Then the properly 
normalized complex null (co-) vector, $(\emm)\:\delta$ will have the form 
(see \refeq{eq:EM} and the definition of $\cQ$, \refeq{Qdef})
\be
\fl \hspace{35pt}\emm =  
\frac{-1}{\rt}\left(\frac{\cR_{,a}\dx{a}}{\cQ}+\rmi \,\cR 
\,\xd\,\varphi\right), \hspace{35pt}\delta =\frac{1}{\rt}\left(\frac{\cQ\,H^{a}}
{H^{c}\cR_{,c}}\ddx{a}+\frac{\rmi}{\cR } \frac{\pd}{\pd \varphi} 
\right),\label{emdelta}
\ee
for some vector $H^{a},$ defined up to an overall scale factor.
 Now, by equations \refeq{eq:D12Q} and  \refeq{eq:d3Q},   
  $\xd\cQ=2\,\cQ (\rho \:\en-\mu \:\el-\tau\,(\emP))$,
so that the co-vector $\rho \,\en-\mu \,\el,\:$ using \refeq{emdelta}, can be written
\be
\rho \:\en-\mu \:\el
=\frac{1}{2\,\cQ}(\cQ_{,a}-\frac{H^{s}\cQ_{,s}}{H^{t}\cR_{,t}}\cR_{,a})\dx{a}.\label{lnMIN}
\ee
Next we observe that the vector $\rho \,\Delta+\mu \,D$ gives zero (by equations 
\refeq{eq:dR} and \refeq{eq:D12Q}) when acting on the 
scalars $\cR,\,\cQ.$ The unique such vector in a 
3-dimensional space is
\be
\rho \:\Delta+\mu \:D =  
\frac{\epsilon^{abc}\,\cR_{,a}\cQ_{,b}}{2\:K}\ddx{c},
\ee
where $K$ is a proportionality factor. 
Then the properly normalized  co-vector  $\rho \,\en+\mu \,\el$ will have the form
\be
\rho \:\en+\mu \:\el= \frac{4\,\rho\,\mu\, K}{\mathbf T\cdot(\nabla\cR \times \nabla\cQ)}
\left(T_{a}-\frac{H^{s}T_{s}}{H^{t}\cR_{,t}}\cR_{,a}\right)\dx{a},\label{lnPLU}
\ee
where the  $T_{a}$ are three arbitrary functions and
\be
\mathbf T\cdot(\nabla\cR \times 
\nabla\cQ)=\epsilon^{abc}\,T_{a}\cR_{,b}\cQ_{,c}.
\ee
Observe that in \refeq{lnPLU} the $T_{a}$ are defined up to an overall 
scale factor and up to the addition of a  multiple of $\cR_{,a}.$
Finally, the vector $\rho \,\Delta-\mu \,D$, which is orthogonal to both  
$\emP$ and $\rho \,\en+\mu \,\el$ and gives $-4\,\mu\,\rho\,\cQ$ 
when acting on $\cQ$, is given by
\be
\rho \,\Delta-\mu \,D=4\,\mu\,\rho\,\cQ\,\frac{\epsilon^{abc}\,\cR_{,a}T_{b}}
{\mathbf T\cdot(\nabla\cR \times \nabla\cQ)} \ddx{c}.
\ee
By equations \refeq{lnMIN} and  \refeq{lnPLU}, the tensor product of 
$\el,\,\en$ depends on the product of the functions $\mu$ and $\rho$ only, 
so the  functions entering the metric are $\cR,\,\cQ,\,H^{a},\,K,\, 
\mu\,\rho,\, T_{a}$ -- a total of 10 functions. Remembering that the  
vectors $H^{a},\, T_{a}$ are 
defined up to scale and that the expressions are invariant if a 
multiple of $\cR_{,a}$ is added to $T_{a}$, we conclude that, in an 
arbitrary coordinate system, the metric 
depends on 7 independent functions, as expected. 

This three-parameter freedom in 
choosing the metric functions $H^{a}, 
\,T_{a}$ can be used to simplify the equations. For example,
 using  equations \refeq{lnMIN} and \refeq{lnPLU}, the closed 2-form 
$\cQ\,\el\wedge\en$ is given by
\be
\cQ\,\el\wedge\en=-\frac{K}{H^{s}\cR_{,s}}\epsilon_{abc}\,H^{a}\,\dx{b}\wedge\dx{c}.
\ee
We can now fix the scale of $H$ by imposing the condition 
$H^{s}\cR_{,s}=K$. Equation \refeq{eq:starFpap} then becomes 
$H^{a}{}_{,a}=0$ -- an equation that can be integrated in terms of a vector potential. 

Collecting everything together, and introducing the index-free notation,
\be
\nabla_{\mathbf H}\cR=H^{a}\cR_{,a},
\hspace{40pt}\mathbf H\cdot \mathbf T=H^{a}T_{a},
\ee
we conclude that, in an arbitrary coordinate basis, the (co-) frame 
vectors have the form
\bea
\rho \:\en+\mu \:\el =4\,\mu\,\rho\,\frac{(\nabla_{\mathbf H}\cR \:T_{a}-
\mathbf H\cdot \mathbf T 
\:\cR_{,a})\dx{a}}{\mathbf T\cdot(\nabla\cR \times \nabla\cQ)},\\
\rho \:\en-\mu \:\el
=\frac{1}{2\,\cQ}(\cQ_{,a}-\frac{\nabla_{\mathbf H}\cQ}{\nabla_{\mathbf 
H}\cR}\cR_{,a})\dx{a}, \\
\hspace{35pt}\emm =  \frac{-1}{\rt}\left(\frac{\cR_{,a}\dx{a}}{\cQ}+\rmi \cR 
\,\xd\,\varphi\right),\\
\rho \:\Delta+\mu \:D =  
\frac{\epsilon^{abc}\,\cR_{,a}\cQ_{,b}}{2\:\nabla_{\mathbf H}\cR}\ddx{c},\\
\rho \:\Delta-\mu \:D = 4\,\mu\,\rho\,\cQ\,\frac{\epsilon^{abc}\,\cR_{,a}T_{b}}
{\mathbf T\cdot(\nabla\cR \times \nabla\cQ)} \ddx{c},\\
\hspace{48pt}\delta =\frac{1}{\rt}\left(\frac{\cQ}{\nabla_{\mathbf 
H}\cR}H^{a}\ddx{a}+\frac{\rmi}{\cR } \frac{\pd}{\pd \varphi} \right).
\eea
These expressions are invariant under a redefinition of the $T_{a}$ 
according to
\be
T_{a} \:\rightarrow \: 
T^{\prime}_{a}=\lambda_{1}\,T_{a}+\lambda_{2}\,\cR_{,a},
\hspace{35pt}\lambda_{1},\,\lambda_{2}\:\mbox{ arbitrary},
\ee
while equation \refeq{R44eq} is satisfied by choosing 
\be
H^{a}=\epsilon^{abc}\mathcal{A}_{b,c}\hspace{40pt}\mbox{for some vector potential }\mathcal{A}_{a}.\label{Apotential}
\ee
Finally, we observe that the linear combinations of the vectors $\el, \en$ appearing 
in equations (60-64) are invariant under the boost freedom \refeq{BoostA}.

\section{Possible Choices of Frame and Coordinate Degrees of Freedom}\label{secBOO}
The expressions for the NP frame vectors given in the previous 
section are invariant under two distinct types of transformation: 
\begin{enumerate}
    \item the remaining freedom in  the choice of unknown functions (the 
ratio $\mu/\rho$ and the arbitrariness in the definition of 
$T_{a}$), and   
    \item  an arbitrary change in coordinates $x^{a}\rightarrow 
   x^{\prime a}(x^{b})$, that can be eliminated by imposing three 
   ``coordinate conditions'' on the 7 independent metric functions.
\end{enumerate}
Ideally, the freedom in the first type of transformation should be used 
to bring the remaining
three ``main''  equations to as simple a form as possible, 
as was done with \refeq{dstrF} and the scaling of $H^{a}$. Then the 
second type of transformation can be used to 
seek solutions of these equations in a system of
coordinates that  
is adapted to the specific physical 
system one is interested in. In that way, it will be easier to 
relate the arbitrary functions of 
integration entering the solution (and the three ``coordinate 
conditions'') to properties  of the physical system.
That the choice of coordinates is important in specifying a 
particular problem is evident from the following consideration:  the ``general'' 
(depending on an arbitrary  function of two variables)
approximate solutions obtained in references \cite{WAYL87} and 
\cite{Bondi62}, using  geometrically defined coordinates, offer no 
clues as to how the arbitrary functions in these solutions are to be 
chosen so as to give 
the metric describing, say,  two coalescing Schwarzschild black holes 
rather than that  of any other distribution of matter along the axis. For 
the two-black-hole 
problem, the  functions of integration would be expected  
to depend ultimately on the properties of the two world lines and 
the two masses 
-- the  physical data needed to specify the problem. These are 
functions of one variable only -- a parameter (proper time) along each world line.  
With this in mind, we have developed and studied 
a coordinate system that uses these two parameters as coordinates 
and is thus appropriate for the two-black-hole problem. 
It will be presented 
in a future publication. For the rest of this section we will discuss 
possible choices of the first kind of transformation.

The freedom in the choice of $T_{a}$ can be used to make the scalars 
$\mathbf T\cdot(\nabla\cR \times \nabla\cQ)$ 
and $\mathbf H\cdot \mathbf T$ equal to anything one pleases. And it 
is not unreasonable to expect that one can exploit this freedom to bring 
one (or more) combination(s) of the equations to a form that can be 
solved. The fact, however, that these scalars involve almost all of 
the free functions in a non-trivial way makes this task extremely 
difficult.

Turning now to the boost freedom \refeq{BoostA}, several choices 
suggest themselves: the simplest 
is to use this freeedom to  set the ratio  $\mu/\rho$ 
equal to unity.  (In fact,  Ludwig \cite{LUD02} uses this choice in 
some cases). 
However, the main equations do not become any 
simpler\footnote{Simplicity is, of course, a subjective criterion. 
What we are really looking for are equations that we know how to 
integrate!} with this choice. Another choice is to use the boost 
freedom \refeq{BoostA}, 
under which   $\alpha+\beta \rightarrow \alpha+\beta + \delta (\ln{A})$, 
to make the sum $\alpha+\beta$ vanish. But this again does not lead to 
equations that we know how to handle.

A more promising  possibility arises in considering the main equation \refeq{main2}.
The effect of a  boost on $G_{0}$,  defined in \refeq{G0}, is
\be 
G_{0}\:\rightarrow\: G_{0}+\frac{1}{2}(\rho\,\Delta+\mu \, D)\ln{A},
\ee
so, in principle, $A$ can be chosen to make $G_{0}$ vanish. Then 
equation \refeq{main2} becomes a 
total divergence, which again can be ``integrated'' in terms of a 
vector potential. Specifically, when $\rho\,\gamma+\mu\,\epsilon=0$, 
equation \refeq{main2} implies that the following 2-form is closed:
\be
\fl\hspace{30pt}F_{2}=\cR  \left[(\mu-2\,\gamma)\,\el\wedge(\emP)+
(\rho-2\,\epsilon)\,\en\wedge(\emP)+2\,\tau\,\el\wedge\en\right].
\ee
However, we have not been able to obtain managable expressions using 
the vector potential arising from $\xd F_{2}=0$ 
together with $\rho\,\gamma+\mu\,\epsilon=0$. 

Finally, the observation that, under the boost \refeq{BoostA} the 
connection one-form 
\be
\omega_{01}=\gamma \,\el+\epsilon\,\en-\alpha\,\emm-\beta\,\emmB
\ee
transforms according to
$\omega_{01} \rightarrow \omega_{01}+\xd A/A$,
suggests that we can impose a Lorentz-type gauge condition on 
$\omega_{01}$
\be
-^{*} \xd(^{*} \omega_{01})=\slashed{D}\gamma+
\slashed{\Delta}\epsilon-\slashed{\delta}(\alpha+\beta)=0,
\ee
which implies that the 2-form
\be
\fl\hspace{30pt}F_{3}=\cR \left[\gamma\,\el\wedge(\emP)-\epsilon\,\en\wedge(\emP)
-(\alpha+\beta)\,\el\wedge\en \right]
\ee
is closed. There are, clearly, many other possibilities. 

\subsection{Properties of Gravitational Principal Null Directions}\label{GPND}
Instead of trying to simplify the general equations as much as 
possible, one can seek solutions that satisfy additional conditions. 
Such conditions can be suggested by examining the properties of 
the Gravitational Principal Null Directions (GPNDs) of the spacetime. 
 For example, equations \refeq{PSIvalues} imply that, in vacuum, the frame vector\ $\el$  
is a  GPND ($\Psi_0=0$) if and only 
if it is geodesic ($\kappa=0$), and the same is true of $\en$. So 
$\kappa=0=\nu$ are 
obvious conditions to use in the search for special solutions.

Consider now an arbitrary null vector in the 3-space orthogonal to 
the Killing trajectories $\el+z^2\,\en +z \,(\emP)$, parametrized by  some 
real parameter $z$. The condition that it is a GPND is
$ \Psi_{0}+4\, \Psi_{1}\,z+6 \,\Psi_{2}\,z^{2}+4 \,\Psi_{3}\,z^{3}+ 
\Psi_{4}\,z^{4}=0$, which, using \refeq{PSIvalues}  in vacuum becomes 
\be
\kappa+6\,\tau\,z^2-\nu\,z^4+4\,(\rho\,z-\mu\,z^3)=0. \nn 
\ee
We now observe that, if we choose $z=\pm\,\sqrt{\rho/\mu}$ (assuming that 
$\rho \mbox { and }\mu$ have the same sign), and require that the 
spin coefficients satisfy the single relation
\be
\kappa \,\mu^2-\nu \,\rho^2+6\,\mu\,\rho\,\tau=0,\label{spinCOND}
\ee
then \emph{both}  null vectors 
\be
N_{\pm}=\mu\,\el+\rho \,\en \pm \sqrt{\mu\,\rho}\,(\emP) \label{Npm}
\ee
will be GPNDs.
 Equation \refeq{spinCOND} is invariant under the boost 
transformation \refeq{BoostA}. Nevertheless, it can be considered as 
determining the ratio $\rho/\mu$ in terms of $\kappa,\,\nu,\,\tau$. 
Then, provided that \refeq{spinCOND} gives two distinct real and positive 
solutions  for this ratio,   equation \refeq{Npm}
will determine altogether four distinct GPNDs. 
These four GPNDs and the frame vectors $\el,\,\en,\,\emm,\,\emmB$ will then be 
related in a way that is analogous to the way 
 the four GPNDs of a Petrov-type I spacetime are related to
the geometrically determined ``Weyl canonical 
frame'' defined in \cite{SB89}. Thus, condition \refeq{spinCOND} on 
the spin coefficients 
can be interpreted as the condition that the ``canonical frame'' 
defined in this paper coincides with the ``Weyl canonical  frame'' 
of  \cite{SB89}. Alternatively, one can impose the condition that one 
of the null vectors \refeq{Npm} is hypersurface orthogonal, say 
$N_{+}\wedge \xd N_{+}=0$, 
thus defining a family of  of Bondi-like null hypersurfaces. 

Note that the extra conditions proposed here do not necessarily imply that 
the Weyl tensor is algebraically special, even though further analysis of the 
equations may lead to that conclusion.

We intend to consider further some of the possibilities suggested here 
in future publications.
\vspace{25pt}

\noindent \emph{Note:}  The results presented in this paper involve extensive calculations. 
All equations given have  been checked using the  computer algebra 
program \textsc{Mathematica} together with the author's package 
``\emph{Exterior Differential Calculus}'', which is available at 
www.inp.demokritos.gr/$\sim$sbonano/EDC/.

\section*{Appendix}
The exterior derivatives of the basis co-vectors are given by (see 
\cite{PandR}, \S 4.13)
\bea
\xd\,\el=(\alpha+\betaB-\tauB) \,
    \el\wedge\emm+(\alphaB+\beta-\tau) \,
    \el\wedge\emmB-(\epsilon+\epsilonB) \,
    \el\wedge\en\\ \nn
    +(\rho-\rhoB) \,\emm\wedge\emmB+
  \kappaB \,\emm\wedge\en+
  \kappa \,\emmB\wedge\en,\\
\xd\,\emm=(\gamma-\gammaB+\muB) \,\el\wedge\emm+
  \lambdaB \,\el\wedge\emmB-(\piB+\tau) \,
    \el\wedge\en\\ \nn
    +(-\alphaB+\beta) \,
    \emm\wedge\emmB+(-\epsilon+\epsilonB+\rho)\,
       \emm\wedge\en+
  \sigma \,\emmB\wedge\en, \\
\xd\,\en=\nu \,\el\wedge\emm+
  \nuB \,\el\wedge\emmB-(\gamma+\gammaB) \,
    \el\wedge\en+(\mu-\muB) \,
    \emm\wedge\emmB\\ \nn
    +(\alpha+\betaB-\pi)\, 
    \emm\wedge\en+(\alphaB+\beta-\piB) \,
    \emmB\wedge\en.
\eea
The assumption that the Killing 1-form is hypersurface orthogonal and 
in the direction $\emM$ 
implies that $(\emM)\wedge\xd(\emM)=0$, which gives the equations
\be
\fl\begin{array}{ccc}
\mu-2 \gamma+\lambda=\muB-2 \gammaB+\lambdaB, &
\hspace{20pt}\rho-2 \epsilon+\sigma=\rhoB-2 \epsilonB+\sigmaB, &
\hspace{20pt}\tau-\pi=\tauB-\piB.\label{sp3eqs}
\end{array}
\ee
In addition, it requires that the exterior derivatives of  $\el, \:\en, \:(\emP)$ must be 
expressible as linear combinations of $\el\wedge\en, \:
\el\wedge(\emP),\:\en\wedge(\emP)$ only. Requiring that the 
coefficients of  $\emm\wedge\emmB, \:
\el\wedge(\emM),\:\en\wedge(\emM)$ vanish, we obtain
\be
\fl\begin{array}{ccc}
\rho=\rhoB, &  \hspace{20pt} \kappa=\kappaB, &
\hspace{20pt} \tau+\alpha-\beta=\tauB+\alphaB-\betaB,\\
  \alpha+\beta=\alphaB+\betaB, &
  \hspace{20pt} \rho-2 \epsilon-\sigma=\rhoB-2 \epsilonB-\sigmaB, &
  \hspace{20pt}\mu-2 \gamma-\lambda=\muB-2 \gammaB-\lambdaB,\\
    \mu=\muB, & \hspace{20pt} \nu=\nuB, & \hspace{20pt} \pi-\alpha+\beta=\piB-\alphaB+\betaB.   
\end{array}\label{sp9eqs}
\ee
The unique solution of the 12 linear equations 
\refeq{sp3eqs}, \refeq{sp9eqs} is that 
all 12  spin coefficients are real.

With all spin coefficients real and the further choice of frame made in 
section \ref{secFR}, for which
$\lambda=\mu, \;\sigma=\rho, \;\pi=-\tau$, the exterior derivatives of 
$\:\el, \:\en, \:(\emP), \:(\emM)$ are given by
\bea
\xd\,\el=(\alpha+\beta-\tau)\, 
    \el\wedge(\emP)-2\,\epsilon\,\el\wedge\en
    +\kappa\, (\emP)\wedge\en, \label{finXDel}\\
\xd\,\en=\nu \,\el\wedge(\emP)
  -2\,\gamma\,\el\wedge\en
    +(\alpha+\beta+\tau) \,(\emP)\wedge\en,\\
\xd\,(\emP)=2\,\mu\, \el\wedge(\emP)
   +2\,\rho \,(\emP)\wedge\en, \\   
\xd\,(\emM)=(\alpha-\beta)\,(\emP)\wedge(\emM).\label{finXDemM}
\eea
These equations are equivalent to the commutators and define the spin 
coefficients when the frame vectors are given in a coordinate basis. 
They are used repeatedly in exterior differentiation of forms.

Acting on functions that do not depend on $\varphi$, the 
three non-trivial commutators are:
\begin{equation*}
[\delta,\, D] = 
(\alpha+\beta+\tau)\,D-2\,\rho\,\delta+\kappa\,\Delta,   
\end{equation*}
\be
[\Delta,\, D] = 2\,\gamma\,D+2\,\epsilon\,\Delta, \label{commut} 
\ee
\begin{equation*}
[\Delta,\, \delta ] = \nu\,D-2\,\mu\,\delta+(\alpha+\beta-\tau)\,\Delta.
\end{equation*}
The equations resulting from the action of these commutators on the scalars $\cR,\:\cQ$ 
are contained  in the Ricci equations obtained in section \ref{redEQU}.

\end{document}